\magnification=\magstep1
\baselineskip=16pt
\hfuzz=6pt

$ $

\vskip 1in

\centerline{\bf On the uncomputability of the spectral gap}

\bigskip

\centerline{Seth Lloyd}

\centerline{Department of Mechanical Engineering, MIT}

\vskip 1cm

\noindent{\it Abstract:} This paper reviews the 1994 proof
that the spectral gap of Hamiltonian quantum systems
capable of universal computation is uncomputable.  

\bigskip

In a recent set of papers [1] Cubitt {\it et al.} prove that the
spectral gap of a Hamiltonian system capable of universal computation
is undecidable: in the subspace of Hilbert space corresponding to
computations that halt, the Hamiltonian has a discrete spectrum
and a finite spectral gap, while in the sector corresponding to
computations that fail to halt, the spectrum is continuous and
there is no gap.     The authors exhibit an elegant planar system
based on aperiodic tilings and show that the spectral gap of
this system is undecidable/uncomputable.    The uncomputability of the
spectral gap is not a new result, however. 
In two papers from 1993 [2] and 1994 [3] I showed
that the spectral gap of Hamiltonian systems capable of universal
computation is uncomputable.  (See also [4].)

The purpose of the current paper is to review the proof of
the uncomputability of the gap.    The systems
investigated in [2-3] are the Benioff [5-8], Deutsch [11],
Feynman [9-10], and Margolus [12] models
for quantum computation: they are less `physicsy' than the system
investigated in [1].    Consequently, the proof of the uncomputability
of the gap is straightforward, as will now be seen.

The fundamental mechanism for the undecidability/uncomputability of the spectral
gap proved in [2-3] is the same as in the more recent work [1].   
The basic intuition is simple.   In computations
that halt, the dynamics explores only a finite region of physical
space.   Accordingly, the Hilbert space explored by the system
is finite-dimensional, the spectrum is discrete and the gap is finite.
In computations that fail to halt, the dynamics explores an
infinite region of physical space.   The Hilbert space is infinite
dimensional, the spectrum is continuous, and there is no gap.

References [2-3] belong to the medieval era of quantum information
theory, pre-Shor and pre-arXiv: the contemporary reader may
compare them to illuminated manuscripts.   I now redescribe
their results in contemporary language.   [2-3] investigated
what were then the only known models of quantum computation [5-12].
In particular,  [2] showed that in such models the spectral decomposition of
a computational state is uncomputable, while
[3] constructed the spectrum and eigenstates
for both unitary [5-8,11] and Hamiltonian [9-10,12] models for quantum 
computation, and shows the uncomputability of
the answer to the question of whether the spectrum
in the eigenspace explored by a computation is discrete and gapped,
or continuous and gapless.      

First, [3] considered unitary models as in Benioff's original
model for quantum computation [5-8] and as in Deutsch's
quantum Turing machine [11].    Let $U$ be the unitary
transformation that advances the operation of the computer
by a single time step, and let $|b_0\rangle$ be the initial
state of the qubits in the computer.   As in Feynman's
quantum computer [9-10] the computer possesses
a clock register with states $|\ell\rangle$. 
The clock register starts in the state $|\ell=0\rangle$ and
is incremented by one for each step of the computation.   
The computation proceeds through a set of orthonormal states,
$$U^\ell |b_0\rangle |\ell=0\rangle = U_\ell U_{\ell-1} \ldots
U_1 |b_0\rangle |\ell\rangle,\eqno(1)$$
where $U_\ell$ is the unitary operator that performs
the $\ell$'th quantum logic operation.   We can write
$$U= \sum_\ell U_\ell \otimes |\ell+1\rangle \langle \ell|.\eqno(2)$$
To ensure unitarity of $U$ we allow $\ell $ to vary from $-\infty$
to $\infty$, and define $U_{-\ell} = U_\ell^\dagger$, so
that for negative values of $\ell$ the computation proceeds
`backwards.'

If the computation halts, we can always set up the computer
so that the time evolution is cyclic [3].   The sign
of the clock state $\ell$ is represented by 
qubit with states $|\pm\rangle$. 
To make halting computations cyclic, when
the halt flag is raised, the sign bit switches from $|+\rangle$
to $|-\rangle$ and the subsequent evolution of the computer
undoes the computation, returning it to its initial state.
If the computation halts after $m/2$ steps, then we have
$$U^m|b_0\rangle |0\rangle = |b_0\rangle |0\rangle.\eqno(3)$$    
Consequently, the eigenvalues of
$U$ are $m$'th roots of unity $e^{2\pi i k/m}$ and the corresponding
 eigenstates are 
$$|k,b_0\rangle = {1\over \sqrt{m}} \sum_{\ell = 0}^{m-1}
e^{-2\pi i k \ell/m} U^\ell |b_0\rangle |0\rangle.\eqno(4)$$
Halting computations correspond to eigenspaces of $U$ 
with a discrete spectrum.
The eigenstates are `plane waves,' uniform superpositions of all the states
in the computation.

By contrast, if the computation
specified by $b_0$ does not halt, then the eigenvalues of 
$U$ are of the form $e^{2\pi i a}$, where $a$ can be any real number
greater than or equal to $0$ and less than $1$. The corresponding
(unnormalized) eigenstates are
$$|a,b_0\rangle =  \sum_{\ell = -\infty}^{\infty}
e^{-2\pi i a \ell } U^\ell |b_0\rangle |0\rangle.\eqno(5)$$
Non-halting computations correspond to eigenspaces of $U$ with
a continuous spectrum.

Feynman transformed the discrete unitary time evolution of quantum
computers into a continuous Hamiltonian time evolution by a simple
trick [9-10].   He added a clock register as above, and looked
at the 
the Hamiltonian 
$$H= U+ U^\dagger.\eqno(6)$$ 
Feynman represented his clock in a `unary' fashion, where
$\ell$ represents the position of a single 1 propagating along
a line of 0's.    With such a clock the Feynman Hamiltonian
can be written as a sum of terms that are at most four-local.

In Feynman's Hamiltonian formulation,
when the computer is prepared in the initial state $|b_0\rangle|\ell=0\rangle$,
the subsequent dynamics governed by the unitary transformation
$e^{-iHt}$ causes the clock
to perform a quantum walk, propagating the computation both
forward and backward in $\ell$.    Because we have assumed $U_{-\ell} = 
U_\ell^\dagger$, both forward and backward propagations
perform the same computation.
Feynman did not treat the distinction between halting and
non-halting programs.   Once one adds the innovation
of reference [3] described above, however, 
so that halting problems correspond to computational dynamics that 
`bounce' back and forth,
while non-halting problems correspond to dynamics
that go on forever,
the spectrum of the Feynman Hamiltonian is uncomputable.

In particular,
the eigenvectors of Feynman's Hamiltonian $H$ are 
the same as those of $U$.   They take the form
the form of equation (4) for halting computations and of equation (5)
for non-halting computations.   The eigenvalues of $H$
are twice the real part of the eigenvalues of $U$: they
take the form $2\cos (2\pi k/m)$ for halting computations,
and $2\cos(2\pi a)$ for non-halting computations.    Halting
computations correspond to eigenspaces with a discrete spectrum 
which has an energy gap.   Non-halting computations correspond to 
eigenspaces with a continuous spectrum which is gapless.
Since the answer to the question of whether a particular
computation halts or not is uncomputable, so is the answer to
the question of whether an initial state lies in an eigenspace
of $H$ with discrete or continuous spectrum.   The gap of $H$
is uncomputable.

The fact that the ground state of the Feynman Hamiltonian
contains all the states in the computation in quantum superposition
underlies the theory of QMA completeness [13] and the equivalence
of adiabatic and conventional quantum computation [14].

In summary, the innovation of Cubitt {\it et al.} [1] is not
to show that Hamiltonian quantum systems capable of universal computation
have an uncomputable gap.    This result was proved in [2-3].
Rather, the contribution of [1] is to present an elegant, two-dimensional,
planar, `physics-like' Hamiltonian system whose overall gap is uncomputable
(as opposed to uncomputable within the eigenspaces corresponding to
different computations, as shown here).
This is a significant accomplishment and as the authors note, suggests
that the gap of other physics-based Hamiltonians might also be 
uncomputable.   Meanwhile, the contemporary reader can return to [2-3, 5-12] 
for the original proof and for a reminder of what life was like in 
the medieval era of quantum information theory.

\vfill\eject

\noindent{\it References:}

\bigskip\noindent [1] T.S. Cubitt, D. Perez-Garcia, M.M. Wolf,
Undecidability of the spectral gap, {\it Nature} {\bf 528},
207-211 (2015); arXiv: 1502.04135, 1502.04573.

\bigskip\noindent [2] S. Lloyd, Quantum computers and uncomputability,
{\it Phys. Rev. Lett.} {\bf 71}, 943-946 (1993).

\bigskip\noindent [3] S. Lloyd, Necessary and sufficient conditions
for quantum computation, {\it J. Mod. Opt.} {\bf 41}(12),
2503-2520 (1994).

\bigskip\noindent [4] S. Lloyd, Uncomputability and physical law, to
appear in {\it The Incomputable,} B. Sanders, ed.;
arXiv: 1312.4456.

\bigskip\noindent [5] P. Benioff, {\it J. Stat. Phys.} {\bf 22}, 563 (1982).

\bigskip\noindent [6] P. Benioff, {\it Phys. Rev. Lett.} {\bf 48},
1581; 1982.

\bigskip\noindent [7] P. Benioff, {\it J. Stat. Phys.} {\bf 29}, 515 (1982).

\bigskip\noindent [8] P. Benioff, {\it Ann. N.Y. Acad. Sci.} {\bf 480},
475 (1986).

\bigskip\noindent [9] R.P. Feynman, {\it Optics News} {\bf 11}, 11 (1985).

\bigskip\noindent [10] R.P. Feynman, {\it Found. Phys.} {\bf 16}, 507 (1986).

\bigskip\noindent [11] D. Deutsch, {\it Proc. Roy. Soc. Lond. A} {\bf 400},
97 (1985); {\it ibid.} {\bf 425}, 73 (1989).

\bigskip\noindent [12] N. Margolus, {\it Ann. N.Y. Acad. Sci.} {\bf 480},
487 (1986).

\bigskip\noindent [13] J. Kempe, A. Kitaev, O. Regev, 
{\it FSTTCS 2004: Foundations of Software Technology and Theoretical 
Computer Science,}
Volume 3328 of the series Lecture Notes in Computer Science, pp. 372-383
(2004); arXiv: quant-ph/0406180.

\bigskip\noindent [14] D. Aharonov, W. van Dam, J. Kempe, Z. Landau, S. Lloyd,
O. Regev,  {\it Proc. 45th FOCS}, 42-51 (2004),
{\it SIAM J. Comp} {\bf 37}, 166-194 (2008); arXiv: quant-ph/0405098. 

\vfill\eject\end